# Effect of substrate-induced strain on transport and magnetic properties of epitaxial $La_{0.66}Sr_{0.33}MnO_3$ thin films

**P. Dey, T. K. Nath [a) and A. Taraphder[†]**

Department of Physics & Meteorology, Indian Institute of Technology, Kharagpur 721302 India

## *Abstract*

Electrical transport and magnetic properties of epitaxial 500 Å $La_{0.67}Sr_{0.33}MnO_3$ (LSMO) thin films, grown on different substrates having different lattice strains, are found to exhibit strong correlation with Jahn-Teller (JT) strain. Our study reveals a sufficiently large JT strain even gives rise to distinct insulating state in LSMO films even below the respective para-ferromagentic Curie temperature, which is a contradiction with the established phase diagram of Sr doped manganites. We have presented a microscopic model for the analysis of our data instead of the usual expansion around the undistorted state generally in use in the literature. The model incorporates two relevant $e_g$ orbitals (and their ordering thereof) and the effect of both JT and bulk strain on the transition temperature via double exchange.

---

[a)]Author for correspondence, E-mail:tnath@phy.iitkgp.ernet.in

[†]also at *Centre for Theoretical Studies, Indian Institute of Technology, Kharagpur 721302 India*

Ever since the discovery of colossal magnetoresistance (CMR) effect in the epitaxial thin film of doped perovskite manganite materials, extensive research have been focused in these materials for possible device applications. CMR thin films exhibit variety of interesting and potentially useful properties including nearly complete spin polarization and very large CMR and thus has attracted considerable attention for magnetic storage technology and also for electronics involving spins. The Jahn-Teller (JT) effect in the $MnO_6$ octahedron is an important ingredient, which presumably gives rise to CMR effect.[1] In fact, the behavior of the manganite thin films depends on the co-operative nature of JT effect, which in turn makes these materials susceptible to strain. Volume-preserving biaxial strain, induced by lattice mismatch between film and substrate, is found to play an important role in tuning electrical-transport, magneto-transport and magnetic properties of film.[2 – 8] In this Letter, our objective is to highlight this fascinating physical phenomenon, i.e., substrate induced strain and its effect on electronic band structure in manganite thin films. In this attempt, we have studied electrical transport and magnetic properties of the canonical double exchange (DE) system $La_{0.67}Sr_{0.33}MnO_3$ (LSMO) manganite high quality epitaxial thin films (500 Å) grown on different substrates having different film-substrate lattice mismatch. Such a mismatch introduces different lattice strains in these films. Our study yield intriguing results exhibiting strong correlation with JT strain. We observed that sufficiently large JT strain can yield a distinct semiconducting/insulating behavior of LSMO films even below the respective $T_C$. This, we believe, is not reported so far in the existing literature and standard phase diagram of Sr doped manganites.[9] We have also presented a model for the analysis of our data of paramagnetic (PM) – ferromagnetic (FM) Curis temperature ($T_c$) vs. strains.

We have grown 500 Å LSMO epitaxial films on four different substrates, namely (001) $LaAlO_3$ (LAO), (001) $SrTiO_3$ (STO), (110) $NdGaO_3$ (NGO) and (001) $La_{0.3}Sr_{0.7}Al_{0.35}Ta_{0.35}O_9$ (LSAT) substrates by $90^0$ off-axis rf-magnetron sputtering using a 2-inch diameter stoichiometric target. An rf (13.56 MHz) power of 100 W was applied. The process was carried out in a flowing gas mixture of argon with oxygen in the 12:8 volumetric ratios with 200 mTorr total partial pressure. The growth temperature for the substrate was optimized at $750^0$ C. After deposition the films were cooled under ambient oxygen pressure of 300 Torr. Surface morphology of the films was studied using high resolution scanning tunneling microscopy (STM) (Digital Instrument). Structural characterization of the films was carried out at room temperature using a four circle x-ray diffractometer with Cu $K_\alpha$ radiation ($\lambda$ = 1.5418 Å). Temperature dependent resistivity ($\rho$) measurements were carried out using a 4 - probe DC technique employing Oxford Instruments' MagLab System 2000™ in the temperature range of 4.2 - 300 K.

Figures 1 (a), (b) and (c) show STM images of LSMO films grown on single crystalline NGO, LSAT and LAO substrates, respectively at a scan area of 1.0 × 1.0 $\mu m^2$. The z-range is approximately 12 nm. LSMO films grown on NGO, LSAT and LAO substrates show spherical, elongated and flat planar connected grains like surface morphology, respectively. These different surface morphologies are due to different lattice strain, induced by different film-substrate lattice mismatch. Details of out-of-plane, in-plane pseudocubic lattice spacing and their corresponding in-plane biaxial strain ($\varepsilon_{xx}$, $\varepsilon_{yy}$) and out-of-plane uniaxial strain ($\varepsilon_{zz}$) for this series of films have been reported earlier.[3] The substrate induced lattice strain can be decomposed into bulk strain, $\varepsilon_B = \varepsilon_{xx} + \varepsilon_{yy} + \varepsilon_{zz}$ and a volume preserving JT strain, $\varepsilon_{JT} = \sqrt{1/6}\,(2\,\varepsilon_{zz} - \varepsilon_{xx} - \varepsilon_{yy})$. We have estimated perovskite unit cell volume ($V_F$) of LSMO films as tabulated in Table 1. The normalized modulation in $V_F$, $V_C^N$ (%) = $(V_B - V_F)/V_B \times 100$, where $V_B$ is the unit cell volume of bulk LSMO, increases with increasing tensile $\varepsilon_B$ [inset in Fig. 2 (a)] thus indicating an expanding $V_F$ with $\varepsilon_B$.

According to the established phase diagram[9] of $La_{1-x}Sr_xMnO_3$ ($0 \leq x \leq 0.6$), for x = 0.33 phase transition occurs from PM metal to FM metal. No insulating state is supposed to appear in bulk LSMO (i.e., x = 0.33) in our working temperature range of 4.2 – 380 K. For convenience, we have tabulated $T_C$ s for this series of LSMO films in Table 1 and also have marked the position of $T_C$ at their corresponding ρ - T curves [Fig. 2]. In case of LSAT and NGO substrates [Figs. 2(a) and (b)], we have obtained the expected metallic behavior of LSMO films even above $T_{C,}$. However, for STO substrate [Fig. 2(c)] peak in ρ vs. T curve seems to flatten above its corresponding $T_C$ without increasing monotonically – a deviation from its expected metallic behavior. Most interestingly, for LAO substrate [Fig. 2(d)] there is a distinct metal/insulator transition at a temperature $T_P$, slightly less than $T_C$ indicating an insulating FM state in a narrow temperature range. Other crucial features are the similar variation of $T_C$ and $T_P$ with $\varepsilon_{JT}$ of this series of LSMO films [insets of Fig. 2 (b) and (d)]. $T_P$ for STO and NGO substrates are derived from a slower varying region of ρ with temperature. This, in turn, implies that fundamentally similar phenomena are governing the electrical transport and magnetic properties of LSMO films.

For optimally doped LSMO DE interaction reaches it maximum. It follows that $T_C$ and $T_P$ s can be scaled simply with the kinetic energy of the carriers i.e., $e_g$ electron hopping or the conduction band width (W) as envisaged by the DE scenario. This makes its magnetic as well as transport properties very sensitive to the JT effect. It is evident that irrespective of sign of $\varepsilon_{JT,}$ both $T_C$ and $T_P$ go down with increasing $\varepsilon_{JT}$ [insets of Fig. 2 (b) and (d)]. Furthermore, our study [Fig. 2] reveals a systematic correlation between gradual deviations from the metallic behavior with increasing $\varepsilon_{JT}$ of LSMO films grown on different substrates. We have attributed these features to the localization of charge carriers due to enhanced JT strain. Quite remarkably, a sufficiently large $\varepsilon_{JT}$, e.g., 3.59 % (for LAO substrate), can bring about insulating behavior even below $T_C$. This is a clear evidence of the localization of carriers due to enhanced JT strain. Furthermore, this insulating state melts (the metallic state is favored) by the application of a magnetic field of H = 5 T [Fig. 2(d)]. This is quite surprising that such strong charge localization effects are offset by a magnetic field as low as 5 T, an indication of competing ground states close by in energy and how easily these systems could be tuned across them. There exists many phase diagrams for $La_{1-x}Sr_xMnO_3$ in terms of x[9], band-width[10] etc. in literature but there is no report of an insulating state in LSMO so far. Here based upon our experimental results we propose a schematic phase diagram for the most canonical DE system LSMO (x ≈ 0.33) with respect to the magnitude of $\varepsilon_{JT}$ [inset in Fig. 2(c)]. The lines are only guide to the eyes. The phase diagram clearly shows that for lower value of $\varepsilon_{JT}$ the behavior of LSMO film is in agreement with the established literature[9], whereas for higher value of $\varepsilon_{JT}$ there is an unexpected emergence of an insulating phase in LSMO.

In fact, effects of $\varepsilon_{JT}$ and $\varepsilon_B$ on $T_C$ and $T_p$ are primarily determined by two competing tendencies. The extension/contraction of bonds due to $\varepsilon_B$ leads to a large reduction/enhancement of the electronic hopping-amplitude. The increased $\varepsilon_{JT}$, on the other hand, renders the effective local potential deeper trapping electrons. It is customary[8] to model these effects purely phenomenologically, in terms of a Taylor expansion around the undistorted state,

$$T_C\left(\varepsilon_b, \varepsilon_{JT}\right) = T_C\left(0,0\right)\left(1 - a\varepsilon_b - b\varepsilon_{JT}^2\right) \qquad \ldots\ldots\ldots\ldots\ldots\ldots\ldots\ldots\ldots\ldots\ldots\ldots\ldots\ldots(1)$$

using two experimental fitting parameters, $a = \frac{1}{T_c}\frac{dT_c}{d\varepsilon_B}$ *and* $b = \frac{1}{T_c}\frac{d^2T_c}{d\varepsilon_{JT}{}^2}$ the first derivative with respect to the even parity JT strain vanishes due to cubic symmetry. While this is a satisfactory method for data analysis, particularly for small distortions, it reveals no underlying physical mechanisms that govern $T_C$ as strain changes as discussed above. We, therefore, present a more microscopic model for the analysis of our data.

In the presence of distortions, the poles of the local Green function for a $MnO_6$ octahedron give the depth of the potential and the JT stabilization energy. In the notation of Kanamori[11], this Green function is obtained easily by,

$$\widetilde{G}_{\alpha\beta}^{-1} = \omega I - \widetilde{H},$$

where, $$\widetilde{H} = -g(Q_3\sigma_3 + Q_2\sigma_1) + \frac{1}{2}KQ^2 \quad \text{.............................................(2)}$$

Here $Q_2$, $Q_3$ are the corresponding normal modes of distortion, $g$ is the local electron-lattice coupling constant and $K$ is the spring contant. $\sigma_1$ and $\sigma_3$ are the Pauli spin matrices and $Q^2 = Q_2^2 + Q_3^2$. $I$ is the $2 \times 2$ identity matrix. If such a system is subjected to a bulk strain, then the increased overlap between this orbital-lattice coupled states of neighbouring octahedra lead to tunneling of the electrons because of the mixing of non-orthogonal states. This effectively reduces the local trapping potential. The transition temperature between two neighbouring Mn ions at $i$ and $i+a$ is, in the conventional DE mechanism, proportional to

$$\frac{1}{2}\sum_{\alpha,\beta} t_{i,i+a}^{\alpha,\beta} a^{-7} [n_{i,\alpha}(1 - n_{1+a,\alpha}) + n_{i,\beta}(1 - n_{1+a,\beta})] \quad \text{..............................(3)}$$

since the hopping takes place via ligand ions[12] ( $t_{Mn\text{-}Mn} \sim V^2_{Mn\text{-}O} \sim (a^{-7/2})^2$ , where $V$ is the Mn 3d -O 2p overlap integral). α, β designate the two $e_g$ orbitals and take values 1, 2 while $n_{i,\alpha}$ is the occupancy of the alpha orbital at the $i$-th site. In a cooperative JT situation that obtains in manganites, the JT distortions are staggered and hence averaging over two orbitals is necessary. Furthermore, if one assumes that the orbital order is strong, the occupancy is overwhelmingly in the lower JT state, the relation for $T_C$ (or $T_P$) could be simply obtained after a straightforward calculation

$$T_C = A\{[\frac{\sqrt{3}}{4}n - \frac{1}{2}n(1-n)]a_x^{-7} - \frac{\sqrt{3}}{4}n + \frac{1}{2}n(1-n)]a_{ij}^{\ 7} - \frac{1}{2}n(1-n)a_z^{-7}\} \quad \ldots\ldots\ldots\ldots\ldots\ldots\ldots(4)$$

where, A is a constant of proportionality. We have used here the Mn – Mn 3d hopping matrix elements $t^{x,y,z}{}_{\alpha,\beta}$ with $|1\rangle = |3d_{x^2-y^2}\rangle$ and $|2\rangle = |3d_{3z^2-r^2}\rangle$ along the three symmetry directions.[13]

Figures 3 (a) and (b) show typical fit of the data using Eq. (4). For comparison, the phenomenological fit is also presented in the same figure using Eq. (1). While this is of course a mere expansion about the undistorted state, our fit, using model Eq. (4) above, is quite satisfactory. We have also carried out same fittings for 250 Å LSMO thin films grown on the same set of substrates [Figs 3(c) and (d)]. For 250 Å films, $\varepsilon_B$ remains tensile for STO, LSAT and NGO substrates, whereas for LAO substrate it becomes compressive. This, in turn, yields expected expansion and compression of $V_F$, which is reflected in its corresponding $V_C^N$ (%) vs. $\varepsilon_B$ plot [upper inset in Fig. 3(c)]. It is evident from Fig. 3 that fits by our model are quite satisfactory, while the fit using Eq. 1, though reasonable for 250 Å films, is quite off the mark for 500 Å films. We have also described the differences between experimental $T_C$ and estimated $T_C$, using both Eq. 1 and 4 [lower insets in Fig. 3]. We have carried out the same fitting for $La_{0.8}Ca_{0.2}MnO_3$ thin films grown on LAO substrate having different values of $\varepsilon_{JT}$ depending upon the thickness of the films.[2] Plot of experimental $T_C$ s with $\varepsilon_{JT}$ [upper inset of Fig. 3(d)] reveals that unlike LSMO films, in this case, $T_C$ is not maximum for the minimum value of $\varepsilon_{JT.}$ This somewhat different dependence of $T_C$ with $\varepsilon_{JT}$ corroborates that two competing tendencies, viz. reduction/enhancement of electronic hopping-amplitude due to change in $\varepsilon_B$ and the localization of conduction electrons due to increased $\varepsilon_{JT}$ are primarily governing $T_C$ s. Quite interestingly, fit to experimental $T_C$ s using our model [upper inset of Fig. 3 (d)] is quite satisfactory and thus gives us a measure of confidence about our model.

In addition, the fit uses as an input the value of the JT trapping potential and therefore can be used to assess the extent of the local orbital ordering. An ordering of nearly 60 – 70 % at the hole ($Sr^{2+}$) doping level of x = 1/3 was necessary to fit the data which indicates a fairly strong local orbital order in the system. Note that due to the staggered nature of the dynamical cooperative ordering, the overall average orbital order in the FM metallic phase is indeed negligible.

In conclusions, our study shows that the electrical and magnetic properties of a canonical DE system like LSMO, where the physics is predominantly dictated by the large $e_g$ bandwidth, are extremely sensitive to modulation of their strain states. We observe that a sufficiently large JT strain may give rise to distinct insulating state in LSMO films, which we believe has not been reported so far in the literature. We have presented a microscopic model for the analysis of our data as well. The conclusions achieved here corroborate our knowledge about manganites - a strongly correlated electronic system having strongly coupled local degrees of freedom – where the tuning of any one degree of freedom induces an appreciable modification in the other degrees of freedom.

# References :

[1] Colossal Magnetoresistive Oxides, edited by Y. Tokura (Gordon and Breach, New York, 2000).

[2] R. A. Rao, D.Lavric T. K. Nath, C.B.Eom, L.Wu and F.Tsui, *Appl. Phys. Lett.* **73**, 3294 (1998).

[3] F.Tsui, M. C. Smoak, T. K. Nath and C.B.Eom, *Appl. Phys. Lett.* **76**, 2421 (2000).

[4] Y. P. Lee, S. Y. Park, Y. H. Hyun, J. B. Kim, V. G. Prokhorov, V. A. Komashko and V. L. Svetchnikov, *Phys. Rev. B* **73**, 224413 (2006) and references therein.

[5] Yan Wu, Y. Suzuki, U.Rudiger, J. Yu, A.D. Kent, T.K. Nath and C.B. Eom, *Appl.Phys. Lett.* **75**, 2295 (1999) and references therein.

[6] A. J. Millis, *Nature* **392**, 147 (1998).

[7] A.B. Shick, *Phys. Rev. B* **60**, 6254 (1999).

[8] A.J. Millis, T. Darling, and A. Migliori, *J. Appl. Phys.* **83**, 1588 (1998); A.J. Millis, A. Goyal, M. Rajeswari, K.Ghosh, R. Shreekala, R. L .Greene, R. Ramesh and T. Venkatesan (unpublished).

[9] A. Urushibara, Y. Moritomo, T. Arima, A. Asamitsu, G. Kido, Y. Tokura, *Phys. Rev. B* **51**, 14103 (1995).

[10]A. Taraphder, *J. Phys. Condens. Matter* **19**, 125218 (2007).

[11] K. Kanamori, *J. Appl. Phys.* 31 (Supplement), 14S (1960).

[12] W. A. Harrison, *Electronic structure and the properties of solids* (Dover, 1989).

[13] T. Maitra and A. Taraphder, *Phys. Rev. B* **68**, 174416 (2003); *Europhys. Lett*t. **65**, 262 (2004).

# Figure Captions :

Figure 1. STM shows different surface topography of 500 Å LSMO films grown on different **(a)** NGO **(b)** LSAT and **(c)** LAO substrates for scan size of $1.0 \times 1.0\ \mu m^2$.

Figure 2. ρ(T) plot of LSMO films at both H = 0 and 5 T grown on **(a)** LSAT, **(b)** NGO, **(c)** STO and **(d)** LAO substrates. Inset in **(a)** shows $V_C^N$ (%) as function of $\varepsilon_B$. Insets in **(b)** and **(b)** show $T_C$ and $T_P$ as a function of $\varepsilon_{JT}$, respectively. Inset in **(c)** shows a schematic phase diagram with respect to the magnitude of $\varepsilon_{JT}$. Lines are only guide to the eyes.

Figure 3. $T_C$ s as a function of **(a)**, **(c)** $\varepsilon_B$ and **(b), (d)** $\varepsilon_{JT}$ for 500 Å and 250 Å LSMO films, respectively. Red filled circle (•) symbol stands for experimental $T_C$, black filled triangle (▲) and filled square (■) symbols stand for $T_C$ s obtained from fits using Eqs. (1) and (4), respectively. Corresponding lower insets show the differences between experimental $T_C$ and estimated $T_C$, using both Eq. 1 (red line and symbol) and Eq. 4 (black line and symbol). Upper inset in **(c)** shows $V_C^N$ (%) as a function of $\varepsilon_B$ of 250 Å LSMO films. Upper inset in **(d)** shows experimental $T_C$ s (black filled square (■) symbol and line) and estimated $T_C$ using Eq. 4 (blue filled circle (•) symbol and line) for $La_{0.8}Ca_{0.2}MnO_3$ thin film as a function of $\varepsilon_{JT}$.

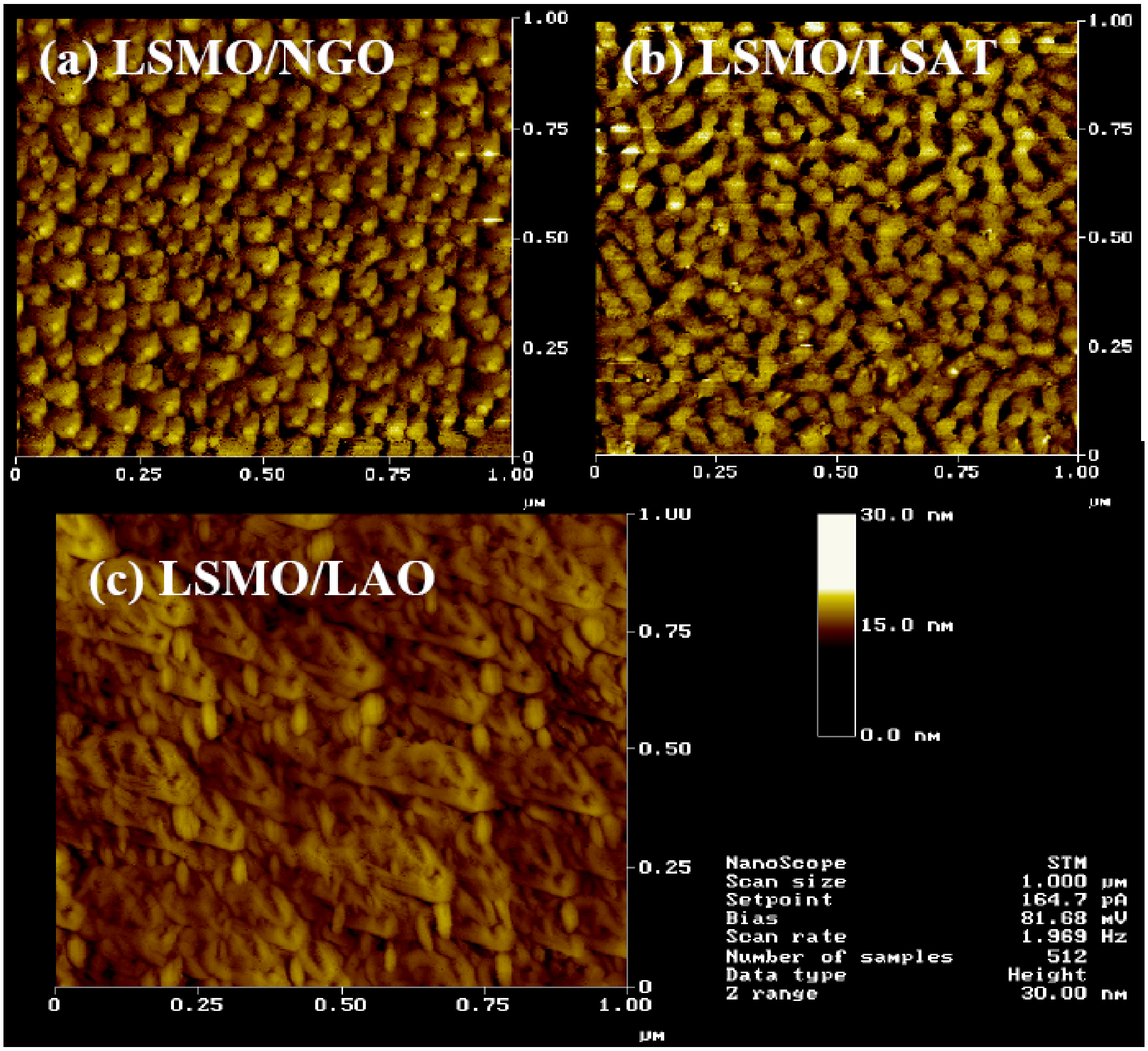

**Fig. 1**

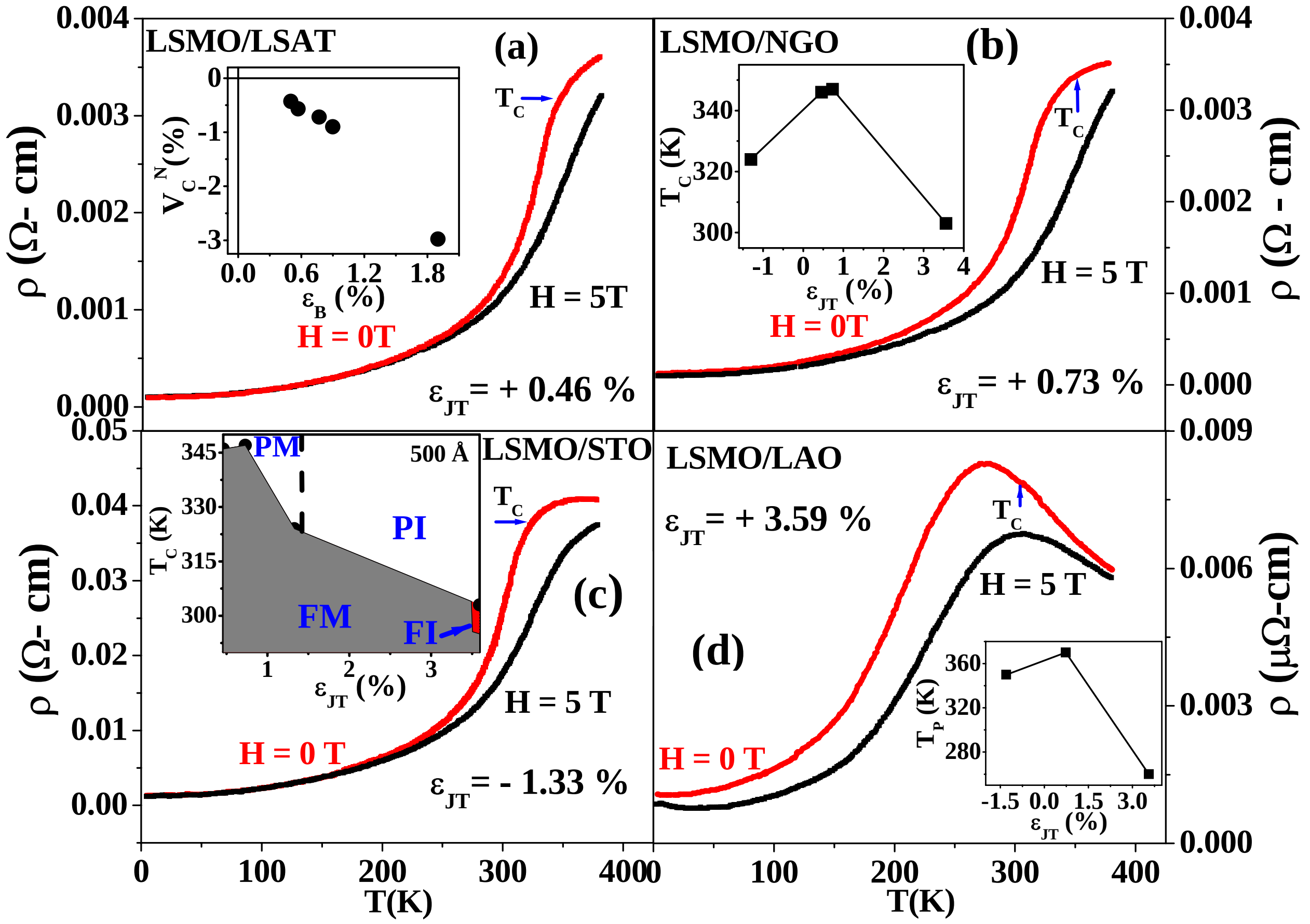
LSMO/LSAT
(a)
$T_C$
H = 5T
H = 0T
$\varepsilon_{JT}$ = + 0.46 %
$V_C^N$ (%)
$\varepsilon_B$ (%)
LSMO/NGO
(b)
$T_C$
H = 5 T
H = 0T
$\varepsilon_{JT}$ = + 0.73 %
$T_C$ (K)
$\varepsilon_{JT}$ (%)
LSMO/STO
(c)
500 Å
PM
PI
FM
FI
$T_C$
H = 5 T
H = 0 T
$\varepsilon_{JT}$ = - 1.33 %
LSMO/LAO
(d)
$\varepsilon_{JT}$ = + 3.59 %
$T_C$
H = 5 T
H = 0 T
$T_P$ (K)
ρ (Ω- cm)
ρ (Ω - cm)
ρ (μΩ-cm)
T(K)

**Fig. 2**

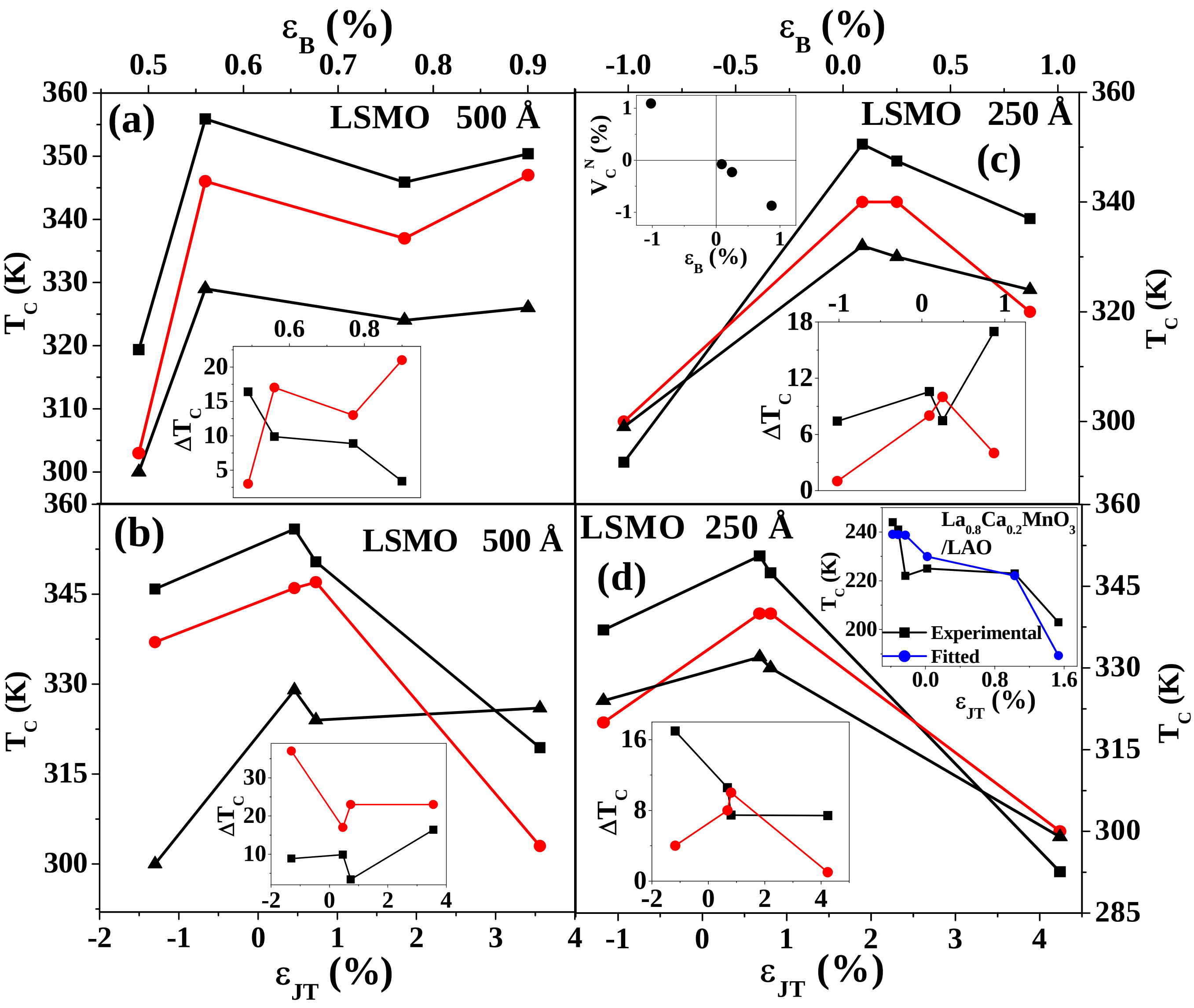

**Fig. 3**